\documentclass{book}    
\usepackage{piers}  
\usepackage{color}

\newcommand{\abs}[1]{\left\lvert#1\right\rvert}

    \DeclareMathAlphabet\mathbfcal{OMS}{cmsy}{b}{n}
	
	\def\U{\rule{0pt}{2ex}U}
	\def\FP{\rule{0pt}{2ex}FP}

	\def\2N{\rule{0pt}{2ex}2N+1-n}

\usepackage[compress]{cite}

\begin{document}

\title{$\mathcal{PT}$-symmetric tight-binding model with asymmetric couplings}
\maketitle
\author      {L. A. M. Rodr\'{i}guez}
\affiliation {Instituto de F\'{i}sica, Benem\'{e}rita Universidad Aut\'{o}noma
  de Puebla}
\address     {}
\city        {Puebla}
\postalcode  {}
\country     {Mexico}
\phone       {}    
\fax         {}    
\email       {rlilian46@gmail.com}  
\misc        { }  
\nomakeauthor

\author      {F. M. Izrailev}
\affiliation {Instituto de F\'{i}sica, Benem\'{e}rita Universidad Aut\'{o}noma
  de Puebla}
\address     {}
\city        {Puebla}
\postalcode  {}
\country     {Mexico}
\phone       {}    
\fax         {}    
\email       {izrailev@ifuap.buap.mx}  
\misc        { }  
\nomakeauthor
\author      {J. A. M\'endez-Berm\'udez}
\affiliation {Instituto de F\'{i}sica, Benem\'{e}rita Universidad Aut\'{o}noma
  de Puebla}
\address     {}
\city        {Puebla}
\postalcode  {}
\country     {Mexico}
\phone       {}    
\fax         {}    
\email       {jmendezb@ifuap.buap.mx}  
\misc        { }  
\nomakeauthor

\begin{authors}

{\bf L. A. Moreno-Rodr\'{i}guez},$^{1}$ {\bf F. M. Izrailev}$,^{1}$ {\bf and J. A. M\'endez-Berm\'udez}$^{1}$\\
\medskip
$^{1}$Instituto de F\'{i}sica, Benem\'erita Universidad Aut\'onoma de Puebla, Apartado Postal J-48, Puebla 72570, Mexico\\

\end{authors}

\begin{paper}

\begin{piersabstract}
We study spectral and transport properties of one-dimensional tight-binding $\mathcal{PT}$-symmetric chains with alternating couplings. Based on the transfer matrix method, we have analytically developed the expressions for the transmission and reflection coefficients for any values of control parameters. These expressions are obtained in a very compact form which separately imbed the generic energy dependence valid for any periodic structure, as well as specific properties of a unit cell composing the scattering setup. Out main interest is in specific properties of the left/right reflections that are due to the $\mathcal{PT}$ symmetric structure of the model. We have found that for the case of asymmetric couplings between dimers, a new type of specific points emerge in the spectrum, which are responsible for quite specific properties of the unidirectional reflectivity. 
\end{piersabstract}

\psection{Introduction}

Optical PT symmetric structures with intermixing balanced gain and loss terms, still attract much attention in view of an experimental possibility to control unusual transport properties of various setups \cite{08Makris,07Ganainy,09Guo,10Rter,09Bendix,08Klaiman}. Among these properties in the first line one has to mention the presence of \textit{exceptional points} \cite{10Longhi,98Bender} in the energy spectra, leading to the so-called \textit{unidirectional reflectivity} \cite{11Lin,11Longhi} for which the reflection from one side of the setup vanishes although the reflection from the other side remains finite.  The onset of this unidirectional reflectivity has been already studied for one-dimensional tight-binding models with symmetric couplings between neighboring sites in the lattice $[\ldots]$ attached to perfect leads. In this type of models it was found that an emergence of the unidirectional reflectivity is directly related to the degeneration of eigenvalues of the transfer matrix which determines all transport characteristics. 

The system we are interested in this paper, consists of a one-dimensional (1D) chain of sites where a particle can hop from one site to nearest-neighbor sites due to non-equal couplings. The symmetric case (constant equal couplings) has already been analyzed in detail \cite{10Longhi}. Here we consider two alternating values for the coupling amplitudes and study how the effect of the coupling asymmetry modifies the transport properties of the chain, obtained by employing the transfer matrix formalism. In Sec. 2 we introduce the model setup and describe spectral properties for the waves propagating through the finite scattering part, attached to perfect leads. In Sec. 3 we apply the transfer matrix formalism and obtain analytical expressions written in the compact form, both for the transmission and reflection coefficients. By analyzing these expressions, we discuss the properties of the transfer matrix in relation to the symmetries present in our PT -symmetric system. After, in Section 4 we explore the transmission and reflection properties of our 1D chain, including the super-transmission zone for two cases that depend on the site-coupling asymmetry. Moreover, we identify the existence of two kinds of transmission resonances. One is similar to that known as the well-known Fabry-Perot resonances, emerging in periodic structure without gain and loss. Another type of resonances are shown to emerge due to specific properties of periodic cells determining the periodic structure of setup.

\psection{The model and energy band structure}
\label{sec:band_structure}

We consider a 1D tight-binding model described by the non-Hermitian Hamiltonian
\begin{eqnarray}
H &=& \sum_{n} \Big \{ \varepsilon_{n} \vert n \rangle \langle n \vert + \nu_{n, n+1} \vert n \rangle \langle n+1 \vert + \nu_{n, n-1} \vert n \rangle \langle n-1 \vert  \Big \}. 
\label{eq:hamiltoniano}
\end{eqnarray}
Here $\varepsilon_{n}$ stands for a site potential corresponding to gain/loss for odd/even $n$,  
\begin{eqnarray}
\varepsilon_{n} &=& (-1)^{n} i \gamma, 
\label{eq:potencial}
\end{eqnarray}
where the balanced gain/loss parameter $\gamma$ is positive, $\gamma>0$. We assume the symmetry $\varepsilon_{n} = \varepsilon^{*}_{2N + 1 - n}$ with respect to the center $n=N$ for a chain of size $2N$. As a result, this model is an example of the $\mathcal{PT}$-symmetric models revealing specific spectral and transport properties.  The case of symmetric couplings, when $\nu_{n, n+1}=\nu_{n, n-1}=\mbox{const}$, has been analyzed widely in Ref.~\cite{15Vazquez}. Below we consider an asymmetric coupling which is characterized by two alternating and non-equal values of $\nu$. Specifically, $\nu_{n, n+1} = \upsilon$ for odd $n$ and $\nu_{n, n-1} = \upsilon$ for even $n$. The 1D chain is attached perfectly to left/right tight-binding perfect leads, see Fig.~\ref{fig:modelo}. Note that this setup can be treated as a 1D chain of $N$ dimers, characterized by an internal parameter $\upsilon$, and coupled to each other by the amplitude $\eta \neq \upsilon$.

\begin{figure}[t!]
\centering
\includegraphics[width=16cm]{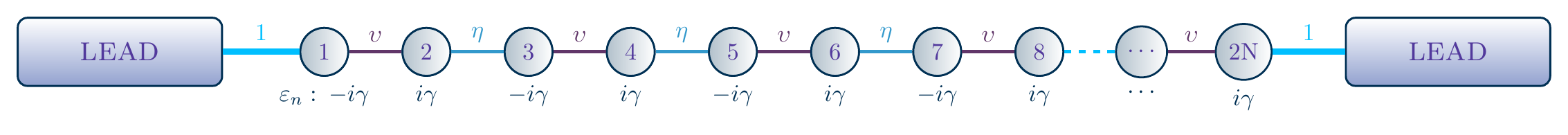}
  \caption{\label{fig:modelo}  {Model setup}} 
 \end{figure}

To find the energy spectra for the waves traveling through the 1D chain, we write the Schr\"{o}dinger equation for the Hamiltonian~(\ref{eq:hamiltoniano}) and look for the solution in the standard form, $\Psi_{n} = e^{-i E t} \: \psi_{n}$. Here $E$ is the energy of the plane wave in the leads, where its wave number $k$ is defined from the relation $E = 2\: \cos(k)$. By substituting $\Psi_{n}$  in the Schr\"{o}dinger equation, we get the set of equations
\begin{eqnarray}
(E + i \gamma) \: \psi_{n} &=& \alpha \psi_{n+1} + \psi_{n-1}, \nonumber \\[1mm]
(E - i \gamma) \: \psi_{n+1} &=& \alpha \psi_{n} + \psi_{n+2} ,
\label{eq:scrhodinger_timeindependent}
\end{eqnarray}
which can be written in the matrix form as follows:
\begin{equation}
\begin{pmatrix}
      \psi_{n+2} \\[1mm]
      \psi_{n+1} 
   \end{pmatrix} = \;
   \begin{pmatrix}
      E - i \gamma \quad & \quad - \alpha \\[1mm]
      1 \quad & \quad 0
   \end{pmatrix} \;
   \begin{pmatrix}
      (E + i \gamma)/\alpha  \quad & \quad - 1/\alpha \\[1mm]
      1 \quad & 0
   \end{pmatrix} \;
     \begin{pmatrix}
        \psi_{n} \\[1mm]
        \psi_{n-1} 
     \end{pmatrix} = \;
    \bf{\mathbfcal{M}} \begin{pmatrix}   
        \psi_{n} \\[1mm]
        \psi_{n-1} 
     \end{pmatrix} 
\label{eq:forma_Matricial} .
\end{equation}
Here we rescaled the energy $E$ and the coupling parameter $\gamma$ as $E \rightarrow E / \eta $ and $\gamma \rightarrow \gamma / \eta $, respectively, and introduced the parameter $\alpha = \upsilon / \eta $. The dispersion relation is defined by the transfer matrix $\mathbfcal{M}$ written for the unit cell (i.e.,~a dimer) of our setup. For this, we compute the eigenvalues of $\mathbfcal{M}$ and write them in the form $\lambda_{1,2} = e^{\pm i 2 \mu}$, where $\mu$ is the wave number for the waves inside the 1D chain. So, by diagonalizing $\mathbfcal{M}$ we find
\begin{equation}
4 \alpha \cos^{2}(\mu) = E^{2} + \gamma^{2} - \delta^{2}\,, \quad \quad E^{2}=4 \cos^{2}(k)\,, \quad \quad  \delta = 1 - \alpha\,.
\label{eq:disp}
\end{equation}

For convenience we defined the asymmetry parameter $\delta=1-\alpha$, which shows how strong is the asymmetry. Note that in the limit $ N  \rightarrow \infty $ the parameter $\mu$ is just the Bloch number. In our model the number of dimers $N$ can be any number, therefore relation~\eqref{eq:disp} can be treated as the dispersion relation for the waves inside the scattering part of the setup, i.e.~the 1D chain. Equation~\eqref{eq:disp} relates the energy $E$ of the transmitted wave to the wave number $\mu$. Moreover, Eq.~\eqref{eq:disp} shows that $\mu$ can be real, complex, or purely imaginary, depending on the model parameters:
\begin{equation}
\mu = 
	\begin{cases} 
      complex  & \mbox{for} \quad E^2 + \gamma^2 < \delta^2 , \\
      real  & \mbox{for } \quad \delta^2 \leq E^2 + \gamma^2 \leq (1+\alpha)^2 , \\  
      imaginary  & \mbox{for} \quad E^2 + \gamma^2 > (1+\alpha)^2 .
   \end{cases}
\label{eq:casos_mu}	
\end{equation}
In the case of symmetric coupling, $\alpha = 1$, the value of $\mu$ is either real or imaginary. Therefore, the asymmetry of couplings results in a new situation with $\mu$ having both real and imaginary parts. 

\begin{figure}[h!]
\centering
     \includegraphics[width=13cm]{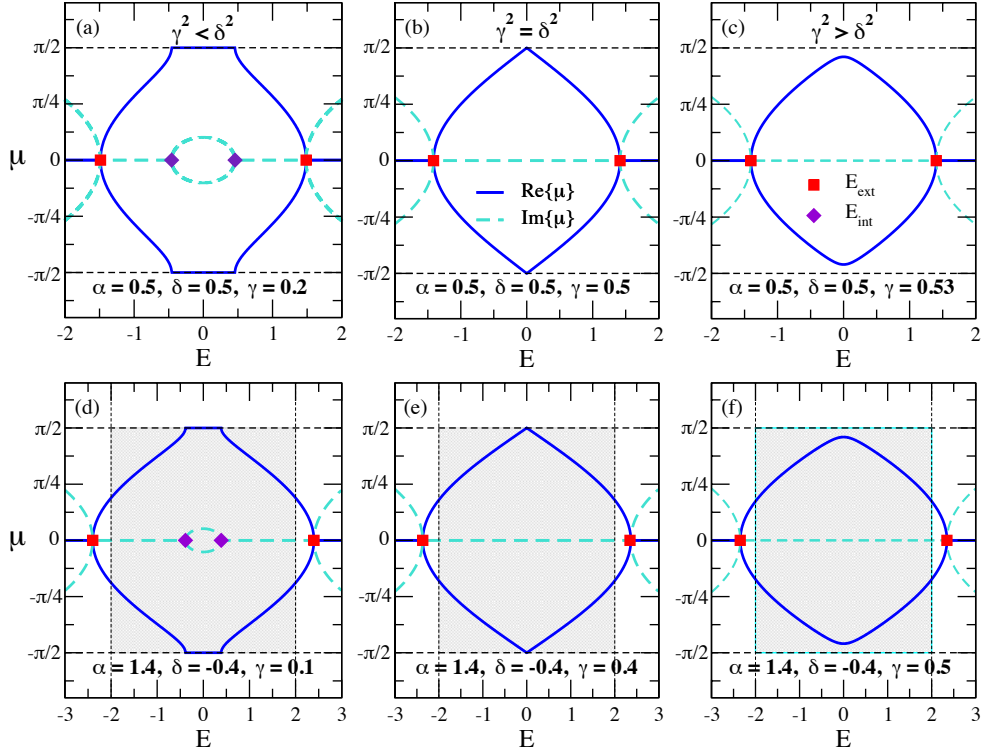} 
     \caption{\label{fig:mu_vs_E}  {Wave number $\mu$ versus energy $E$ for (a-c) $\alpha<1$ and (d-f) $\alpha>1$. Solid and dashed curves correspond to real and imaginary parts of $\mu$, respectively.  Three cases are shown: (a,d) $\gamma^2 < \delta^2$, (b,e) $\gamma^2 = \delta^2$, and (c,f) $\gamma^2 > \delta^2$. Full squares and full diamonds stand for exceptional points $E_{ext}$ and $E_{int}$, respectively, see the text. } }
\end{figure}

The global properties of the dispersion relation are shown in Fig.~\ref{fig:mu_vs_E} for the energies inside the band $E \leq |2|$. For the energies outside the band, the wave number $\mu$ is always imaginary, as it happens for a perfect structure, $\gamma =0$ and $\alpha=1$. It is also known that for the symmetric case, $\gamma > 0$ and $\alpha=1$, the {\it external band edges}, $E_{ext}$, are shifted towards the band center (squares in Fig.~\ref{fig:mu_vs_E}). The same happens for the symmetric case, $\gamma > 0$ and $\alpha \neq 1$. The value of $E_{ext}$ can be easily obtained from Eq.~(\ref{eq:casos_mu}) as
\begin{equation}
E_{ext}^{2} = (1 + \alpha)^{2} - \gamma^{2} = (2-\delta)^{2} - \gamma^{2} .
\label{eq:bordes_externos}
\end{equation}
At this value, the two eigenvalues merge $\lambda_{\pm}=1$ since the wave number vanishes, $\mu=0$. However, in our case of asymmetric couplings, there is another energy value for which 
the eigenvalues are also equal, $\lambda_{\pm}=-1$; this occurs at $\mu=\pi/2$. The corresponding {\it internal band edges}, $E_{int}$, are shown in Fig.~\ref{fig:mu_vs_E} as diamonds. The values of $E_{int}$ are given by
\begin{equation}
E^{2}_{int} = (1 - \alpha)^{2} - \gamma^{2}=\delta^{2} - \gamma^{2} .
\label{eq:bordes_internos}
\end{equation}
Thus, we have two different types of {\it exceptional points}, $E_{ext}$ and $E_{int}$. As is seen from Eq.\eqref{eq:bordes_externos}, for small $\gamma$ and symmetric coupling, $\alpha=1$, the exceptional points $E_{ext}$ are very close to the band edges of the energy, where the reflections vanish. However, we will show below that for the asymmetric case, $\alpha \neq 1$, the reflections vanish for different energy values, termed in this paper as the {\it U-points} which are close to the center of the energy band. These points can be identified from the exact expressions of the reflection coefficient, as we show below.

\psection{Transfer matrix structure}
\label{sec:transfer matrix}

In order to describe transport properties for the model shown in Fig.~\ref{fig:modelo} we follow the transfer matrix approach described in \cite{08Markos}. Note that our final expression for the total transfer matrix $\mathbf{M}$ connecting left and right ends of the scattering set up, is obtained in a compact form allowing to relate the properties of the energy spectrum to the transport properties in a transparent way. To do this, first, we have to obtain the product of $N$ matrices $\mathbfcal{M}$ and present this product in the plane wave representation,
\begin{eqnarray}
{ \bf{\mathbfcal{M}}}^{N} &=& {\bf\mathbf{V}} \left( {\bf\mathbf{V}}^{-1} { \bf{\mathbfcal{M}}} {\bf\mathbf{V}} \right)^{N} {\bf\mathbf{V}}^{-1} \, ,
\label{eq:multi_matrices}
\end{eqnarray}
where 
\begin{eqnarray}
\begin{aligned}[b]
   {\bf\mathbf{V}} = \;
     \begin{pmatrix}   
       {\displaystyle \frac{E - i \gamma}{1+ \alpha \lambda_{2} } } & {\displaystyle \frac{E - i \gamma}{1+ \alpha \lambda_{1} } }\\[3mm]
        1 & 1
     \end{pmatrix},   
\end{aligned} 
\:
\: \: \: \:  
\begin{gathered}[t]
{\bf\mathbf{V}}^{-1} = \frac{1}{2 \alpha \sin(2 \mu)} \;
     \begin{pmatrix}   
        - (E + i \gamma) & - i (1 + \alpha \lambda_{2}) \\[1mm]
         (E + i \gamma) & i (1 + \alpha \lambda_{1})
     \end{pmatrix}   
\end{gathered} ,
\label{eq:eigenvectors}     
\end{eqnarray}
with $\lambda_{1,2} = e^{\pm i 2 \mu}$. Then, we need to construct 
\begin{equation}
{\bf{M}}= {\bf\mathbf{Q}}^{-1} { \bf{\mathbfcal{M}}}^{N} \bf\mathbf{Q} \, ,
\label{eq:matriz_transferencia_completa}
\end{equation}\\
where the matrix $\bf\mathbf{Q}$ takes into account the coupling of the scattering region with the left/right perfect leads,
\begin{equation} 
    \bf\mathbf{Q} =  \;
   \begin{pmatrix}
      1 & 1 \\[1mm]
      e^{- i k} & e^{i k}
   \end{pmatrix},
\qquad 
\begin{gathered}[t]
{\bf\mathbf{Q}}^{-1} =  \frac{1}{2 \: i \sin (k)} \;
   \begin{pmatrix}
      e^{i k} & -1 \\[1mm]
      -e^{- i k} & 1
   \end{pmatrix}
\end{gathered} .
\end{equation}
As a result of a cumbersome analytical procedure we have found the way to get a quite compact and useful form for the matrix elements of the total transfer matrix $\bf{M}$:
\begin{align}
{\bf{M}}_{11} &=  \cos(2 N \mu) + i  \:  \frac{ \sin(2 N \mu) }{ \sin(2 \mu)} \: \left[ \frac{1}{\alpha} \sin(2k) + \frac{3}{2} \cos(k)  \bigg( F_{+} + F_{-} \bigg)  \right] \nonumber \, , \\[1mm] 
{\bf{M}}_{12} &= i e^{ik} \frac{\sin(2 N \mu)}{\sin(2 \mu)} F_{+}(\gamma, \alpha, k) \, , \nonumber \\[1mm]
{\bf{M}}_{21} &= -i e^{-ik} \frac{\sin(2 N \mu)}{\sin(2 \mu)} F_{-}(\gamma, \alpha, k) \, , \nonumber \\[1mm]
{\bf{M}}_{22} &= \cos(2 N \mu) - i  \:  \frac{ \sin(2 N \mu) }{ \sin(2 \mu)} \: \left[ \frac{1}{\alpha} \sin(2k) + \frac{3}{2} \cos(k)  \bigg( F_{+} + F_{-} \bigg)  \right] \, .
\label{eq:entries_matrixMT}
\end{align}

From the analysis of the structure of these matrix elements, first we would like to note that all expressions contain the term $ G(N,\mu) = \sin(2 N \mu) /  \sin(2 \mu)$, which explicitly depends on the size $2N$ of the 1D chain and on the wave number $\mu$. This term is typical for any periodic setup and all details of the unit cell which depend on the control parameters $\gamma$ and $\alpha$ (apart from the energy $E$ which enters through $\mu$ and $k$) are separated from $G(N,\mu)$. A distinctive 
feature of the above expressions is their dependence of the real functions $ F_{+}$ and $ F_{+} $. These functions have the following form: 
\begin{equation}
F_{\pm}(\gamma, \alpha, k) = \frac{\pm \gamma \sin(k) + \frac{1}{2} (\alpha^{2} - \gamma^{2} - 1)}{\alpha \sin(k)}.
\label{eq:funcion_F}
\end{equation}

The representation of the matrix elements of $\bf{M}$ given above allows to understand easily some of the global peculiarities of the scattering. For example, the condition $G(N,\mu) = 0$ determines the resonance value of $E$ for which the transmission through the total structure of size $2N$ is perfect, $T=1$. In the absence of gain and loss, these resonances are known as the {\it Fabry-Perot resonances}. Another property, namely, the vanishing of the reflection coefficients (left or write reflection) is defined by the vanishing of either functions $F_{+}$ and $F_{-}$ (see below). Therefore, the specific energy values for which the transmission is zero can be found just by analyzing the structure of these functions. The presentation of the matrix $\bf{M}$ containing the functions $F_{\pm}(\gamma, \alpha, k)$ (however, different ones) has also been suggested in \cite{13Ramirez} for another $\mathcal{PT}$-symmetric model composed by two layers with the balanced gain/loss structure, coupled non-perfectly to leads.

The following expressions can be useful when analyzing the properties of $F_{\pm}(\gamma, \alpha, k)$:
\begin{eqnarray*}
 F_{+}  F_{-}  = \frac{1}{\alpha^{2} \: \sin^{2}k} \left[ \Big( \alpha \cos(2 \mu) - \cos(2k) \Big)^{2} - \gamma^{2} \sin^{2} k   \right] \, . \nonumber
\end{eqnarray*}
With the use of the relation $F_{+} - F_{-} = 2 \: \gamma/\alpha$ we can also write
\begin{eqnarray}
\nonumber
\alpha \cos(2 \mu) - \cos(2k) = \frac{1}{2} \:  \alpha \: \sin(k)  \Big( F_{+} + F_{-} \Big) \, .
\label{eq:suma_de_Fs}
\end{eqnarray}
From expressions~(\ref{eq:entries_matrixMT}) we can reveal a specific symmetry of the transfer matrix $\bf{M}$:
\begin{eqnarray}
{\bf{M}}_{22}={\bf{M}}_{11}^{*}, \: \: \: \: \: {\bf{M}}_{21}(\gamma)={\bf{M}}_{12}^{*}(- \gamma), \: \: \: \: \: \mbox{Re}\left[ {\bf{M}}_{12}\right] =\mbox{Re}\left[ {\bf{M}}_{12}\right] \equiv 0 \, ;
\label{eq:complex_cunjugate}
\end{eqnarray} 
together with the general relation
\begin{eqnarray}
 \nonumber 
\det ({\bf{M}}) = {\bf{M}}_{11} {\bf{M}}_{22} - {\bf{M}}_{12} {\bf{M}}_{21} = 1.
\label{eq:determinante1}
\end{eqnarray}
The latter relation is typical for transfer matrices, however, the symmetric properties of the matrix $\bf{M}$ are unusual, see Eq.~(\ref{eq:complex_cunjugate}), as compared with those related to $\mathcal{PT}$-symmetric Hamiltonians~\cite{13Mostafazadeh,10Longhi}. Indeed, a typical symmetry of the $\mathcal{PT}$-symmetric transfer matrices corresponds to $\bf{M}_{22}={\bf{M}}_{11}^{*}$ and ${\bf{M}}_{21}(\gamma)={\bf{M}}_{12}^{*}(\gamma)$. As one can see, the second relation is different from that in Eq.~(\ref{eq:complex_cunjugate}). This fact stresses that the $\mathcal{PT}$-symmetry of the Hamiltonian is not uniquely related to the symmetry of the transfer matrix, see discussion in~\cite{17Ramirez}. As for the first relation, $\bf{M}_{22}={\bf{M}}_{11}^{*}$, it corresponds to the pseudo-unitary symmetry which is common in several $\mathcal{PT}$-symmetric Hamiltonians ~\cite{13Mostafazadeh,10Longhi,12Li,13Yu} (see also the discussion in~\cite{13Ramirez}). 

\psection{Transport properties}
\label{sec:transfer matrix2}

Following the standard approach of the scattering theory (see for example~\cite{08Markos}) we can express the transmission coefficient $T$ and the left/right reflection coefficients $R_L$/$R_R$ as 
\begin{equation}
T =  \left[ 1 + {\bf{M}}_{12} {\bf{M}}_{21} + {\bf{M}}_{22} \left( {\bf{M}}_{22}^{*} {\bf{M}}_{11} \right)  \right]^{-1} =\abs{ {{\bf{M}}_{22}}}^{-2},
\qquad
\frac{R_{L}}{T} =  \abs{ {\bf{M}}_{21} }^{2},	
\qquad
\frac{R_{R}}{T} = \abs{{\bf{M}}_{12}}^{2} .
\label{eq:def_trans_refle_terminos_de_entradas}
\end{equation}

From Eqs.~(\ref{eq:def_trans_refle_terminos_de_entradas}) and taking into account that $\det({\bf{M}})=1$, a relation between the transmission and reflection coefficients reads
\begin{equation}
\abs{ 1 -T} = \sqrt{R_{R}R_{L}} \, .
\label{eq:relacion_conservacion}
\end{equation}
It can be seen that the conventional relation $T+R=1$ is recovered in the case when ${\bf{M}}_{21}={\bf{M}}_{12}^{*}$ only. Moreover, this happens when $F_{+}=F_{-}$, therefore for $\gamma=0$, see Eqs.~(\ref{eq:entries_matrixMT}). In this case $R_R=R_L$ and $T \leq 1$ for any energy $E$. If $\gamma > 0$, the transmission coefficient can be larger than one, $T>1$, which results in the relation $T - \sqrt{R_{R}R_{L}} = 1$. Also, note that it is possible to have $R_R=R_L+R$, therefore, $T-R=1$.  


\psubsection{Transmission}
\label{sec:transmission}

The transmission coefficient $T$ can be obtained via the matrix element ${\bf{M}}_{22}$, see Eqs.~(\ref{eq:entries_matrixMT}) and~(\ref{eq:def_trans_refle_terminos_de_entradas}). It can be written in the following form: 
\begin{figure}[h!]  
 \centering
    \includegraphics[width=7.5cm]{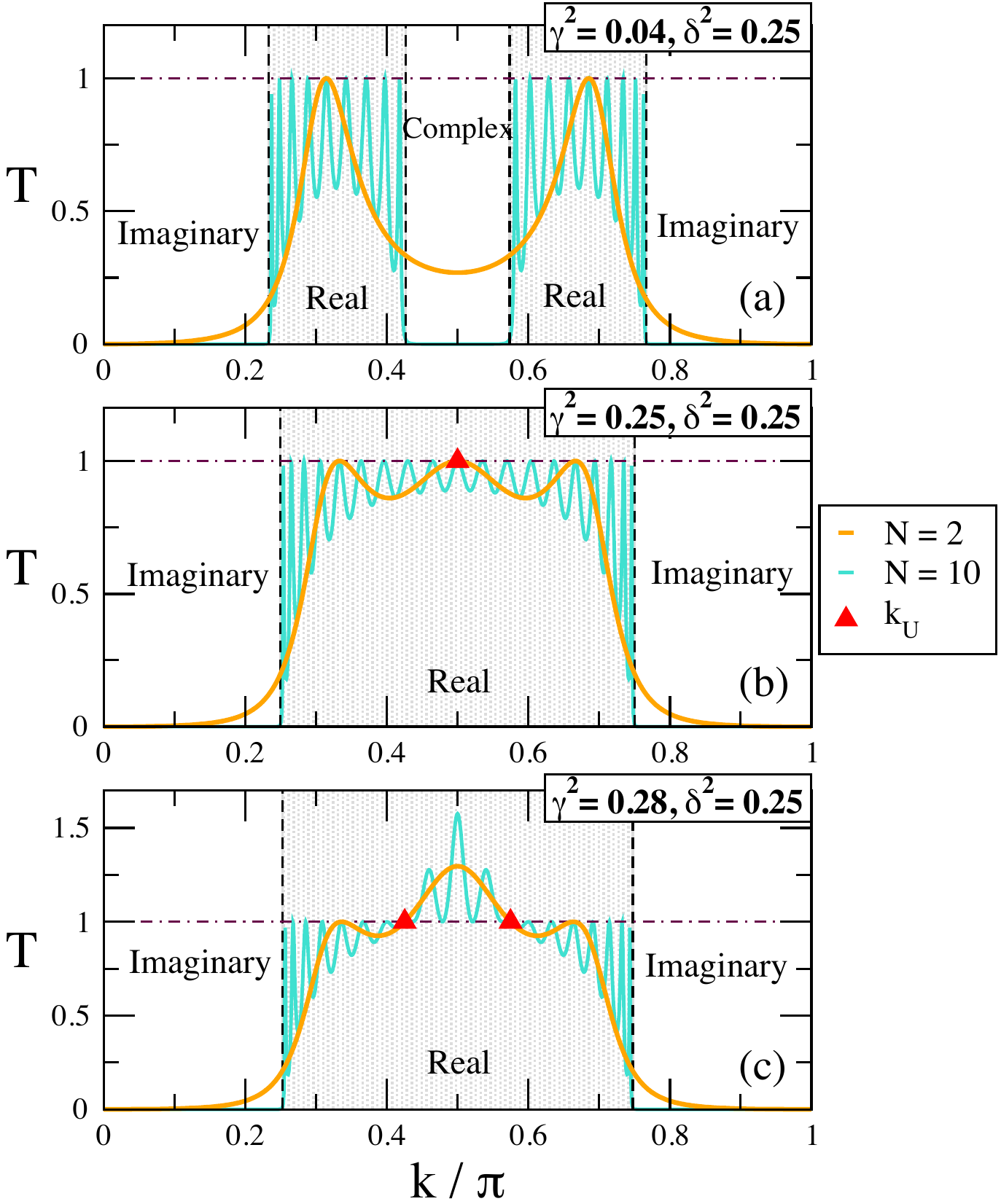}\hspace{0.5cm}
    \includegraphics[width=4.5cm]{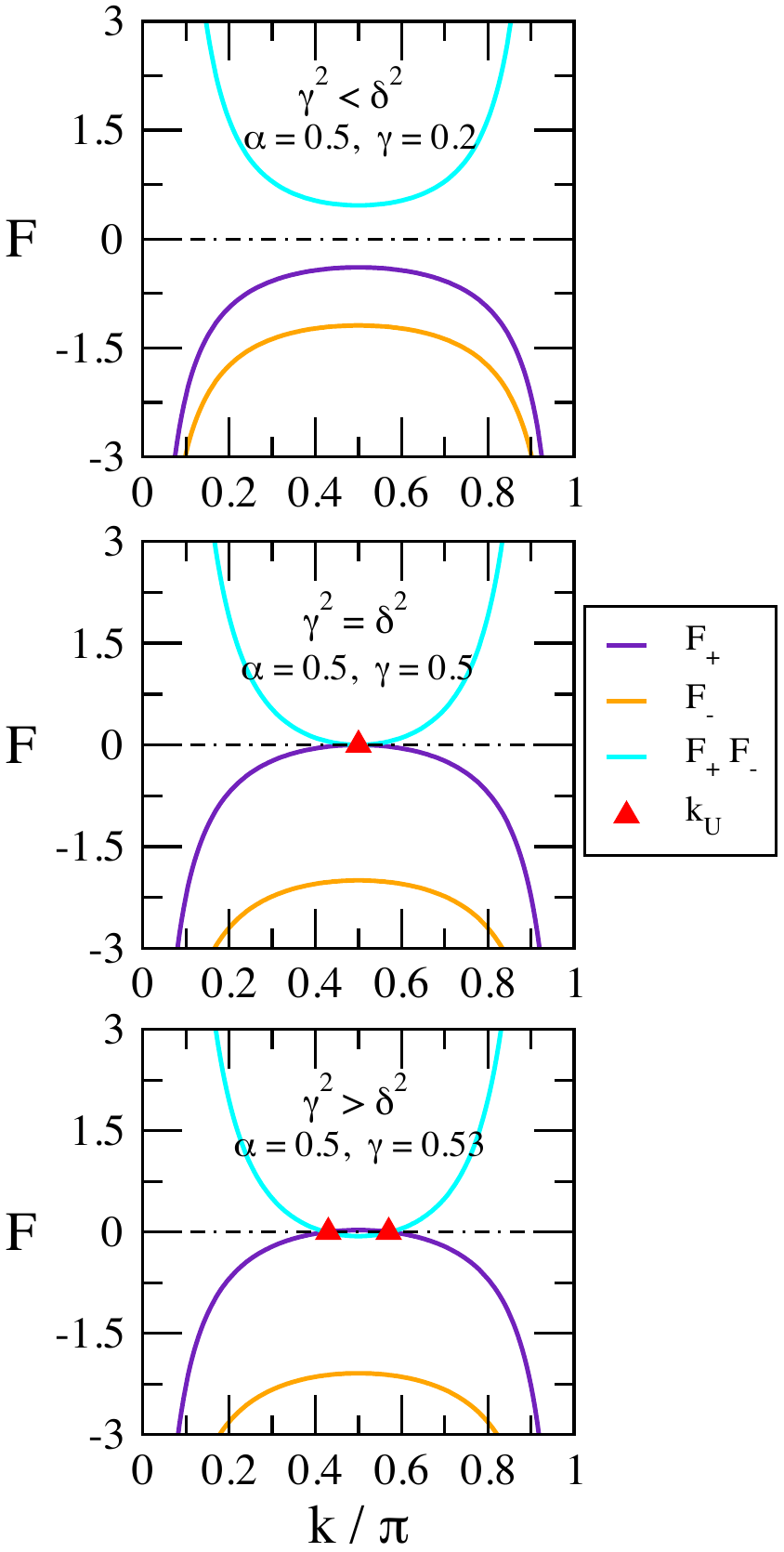}
 \caption{ {Left panel: Transmission coefficient versus $k/\pi$ for the parameters of Fig.~\ref{fig:mu_vs_E}(a-c)  and different number of unit cells, $N=2$ and $N=10$. The shaded regions correspond to real values of $\mu$, marked by external borders $k_{ext}$ (in all cases) and additional internal borders $k_{int}$, according to Eqs.~(\ref{eq:bordes_externos}) and~(\ref{eq:bordes_internos}), for case (a). The triangles stand for specific $U$-points $k_{\U}$ at which one of the $F_{\pm}$ functions vanishes. Right panels: Functions $F_{\pm}$ and their product in correspondence with left panel; one can see that only the function $F_{+}$ can vanish, apart from the special case $\gamma^{2} = \delta^{2}$ at the band center, $E=0$. }}
 \label{fig:T_N2_10_y_F}
\end{figure}

\begin{equation}
T  =  \frac{1}{1 + {\displaystyle \frac{\sin^2(2 N \mu)}{\sin^2(2 \mu)}} F_{+}(\gamma, \alpha, k) F_{-}(-\gamma, \alpha, k)} \, .
\label{eq:transmision}
\end{equation}

The dependence of $T$ versus the wave vector $k$ (according to $2 \cos k = E$) is shown in 
Fig.~\ref{fig:T_N2_10_y_F} for the parameters of Fig.~\ref{fig:mu_vs_E} with $N=2$ and $N=10$. One can see that for $\gamma <  \abs{ \delta} = 1-\alpha$, see Fig.~\ref{fig:mu_vs_E}(a), an additional energy window emerges around the band center $E=0$ (or, equivalently, around $k=\pi/2$). This result is entirely due to the asymmetry of the coupling between nearest sites. When $\gamma^2 =\delta^2$, see Fig.~\ref{fig:mu_vs_E}(b), this window disappears. The data clearly demonstrates that by increasing $N$ the borders between real and complex values of $\mu$ are getting sharper, thus creating true band edges in the limit $N \rightarrow \infty$.

\begin{figure}[h!]  
 \centering
     \includegraphics[width=7.5cm]{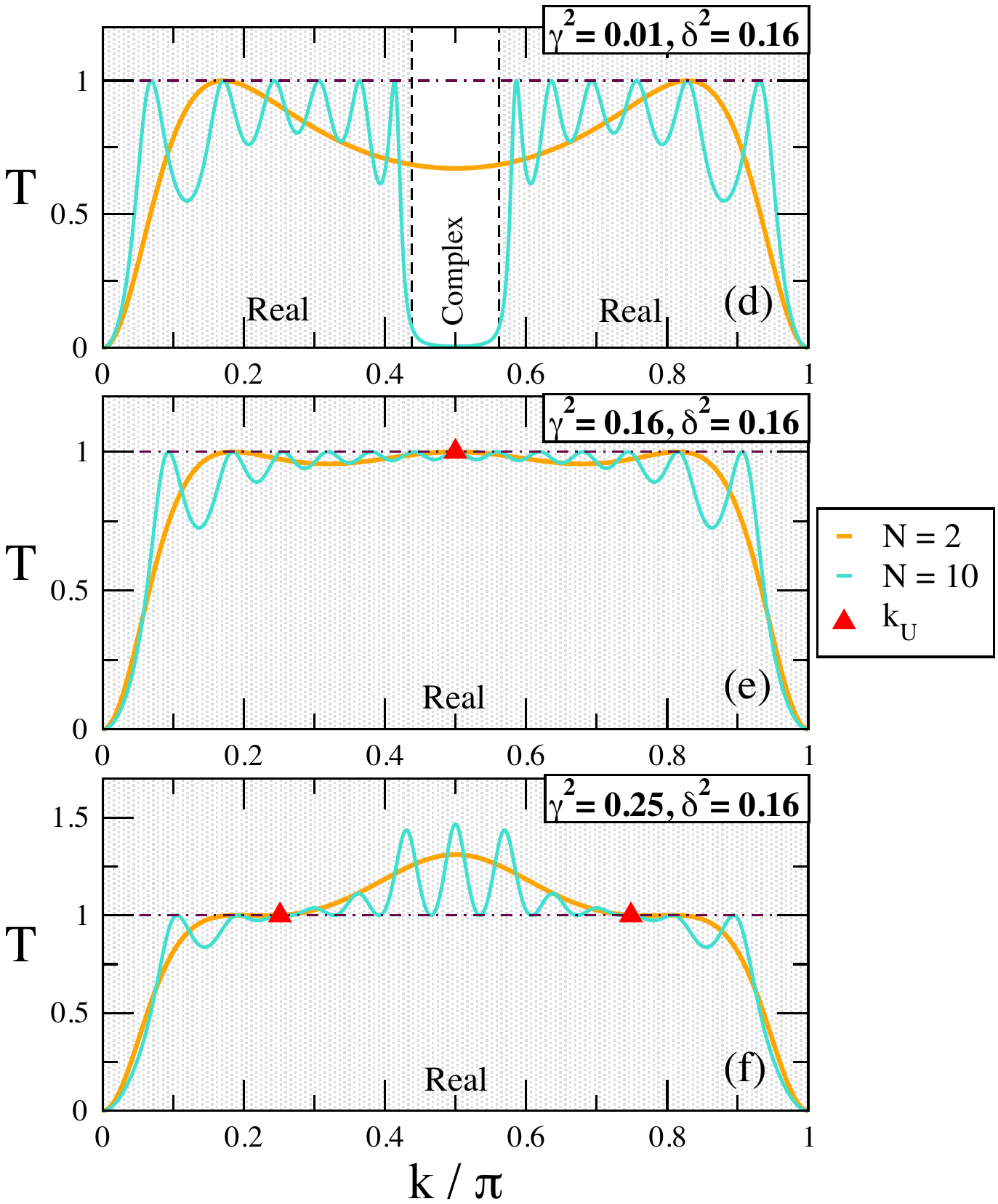}     \hspace{0.5cm}
       \includegraphics[width=4.47cm]{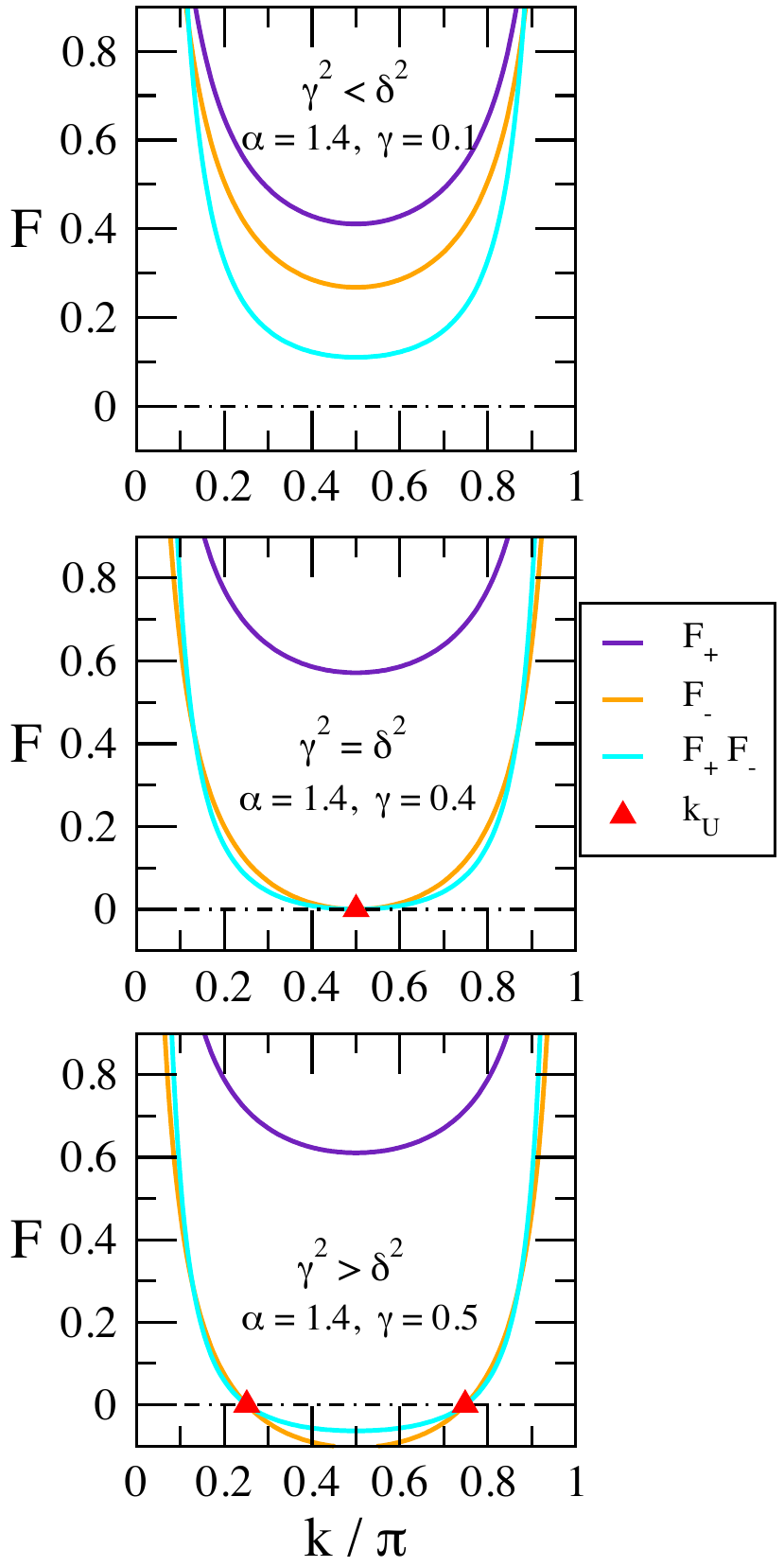}
 \caption{ {Left panel: Transmission coefficient versus $k/\pi$ for the parameters of Fig.~\ref{fig:mu_vs_E} (d-f) and different number of unit sells, $N=2$ and $N=10$. The shaded regions correspond to real values of $\mu $, marked by external borders $k_{ext}$ (for all cases) and additional internal borders $k_{int}$, according to Eqs.~(\ref{eq:bordes_externos}) and~(\ref{eq:bordes_internos}) for the case (a). The triangles stand for specific $U-$points $k_{\U}$ for which one of the $F_{\pm}$ functions vanishes. Right panel: Functions $F_{\pm}$ and their product in correspondence with left panel; one can see that only the function $F_{-}$ can vanish, apart from special case $\gamma^{2} = \delta^{2}$ for the band center, $E=0$.}}
 \label{fig:T_N2_10_y_F-}
\end{figure}
As we already noted, the perfect transmission, $T=1$, emerges in two cases: either due to the Fabry-Perot resonances defined by the vanishing of the term $ G(N,\mu) = \sin(2 N \mu) /  \sin(2 \mu)$, or by the zeros of the functions $F_{\pm}(\gamma, \alpha, k)$. The resonance energies $E_{FP}$ corresponding to the Fabry-Perot resonances can be obtained from the relation, $\mu = m \pi/2 N, \: \: \: \: m = 1, \ldots, N-1$. With the use of the dispersion relation (\ref{eq:disp}) one gets,
\begin{equation}
E_{\FP}^{2} = 4 \: \alpha \: \cos^2 \left( \frac{m\pi}{2N} \right) - \gamma^{2} + \delta^{2} ; \qquad 2 \cos k_{\FP} = E_{\FP} .
\label{eq:e_fp_g<d}
\end{equation} 
Another option for $T=1$ is due to vanishing one of the functions $F_{\pm}$, as it stems from Eq.\eqref{eq:transmision}. The analysis of these functions shows that for the parameters chosen, only one of the functions $F_{\pm}$ can vanish either for two energy values (when $\gamma ^{2} > \delta ^{2}$) or at the band center (when $\gamma ^{2}=\delta ^{2}$) depending on the $\alpha$ parameter. Fig.  \ref{fig:T_N2_10_y_F} shows that $F_{+}$ vanishes for $\alpha < 1$, and $F_{-}$ vanishes for $\alpha > 1$. This fact is seen in Fig. \ref{fig:T_N2_10_y_F-}. In the latter case the both functions vanish together; this case is very specific and will be discussed below in more detail. The specific $U-$ points where $F_{\pm}=0$ for $\gamma ^{2} > \delta ^{2}$ are defined by the relation,
\begin{equation}
E^{2}_{\U} = \frac{1}{\gamma^{2}} \left[ \gamma^{2} - \delta^{2} \right] \left[ \left( 1 + \alpha^{2} \right) - \gamma^{2} \right] 
\label{eq:resonancias_U_depE}
\end{equation}
These points are marked by triangles in Fig.\ref{fig:T_N2_10_y_F}. 


\psubsection{Reflection}
\label{sec:reflection}

According to Eqs.~(\ref{eq:def_trans_refle_terminos_de_entradas}) we can represent the expressions for left and right reflection coefficients, $R_L$ and $R_R$, respectively, in the following compact form
\begin{eqnarray}
\frac{R_{L}}{T}  = \frac{\sin^2(2 N \mu)}{\sin^2(2 \mu)} F_{-}^{2}(-\gamma, \alpha, k),  & \quad  \quad &  \frac{R_{R}}{T}  = \frac{\sin^2(2 N \mu)}{\sin^2(2 \mu)} F_{+}^{2}(\gamma, \alpha, k).
\label{eq:reflexiones}
\end{eqnarray}

\begin{figure}[h!]
   \centering
     \includegraphics[width=14cm]{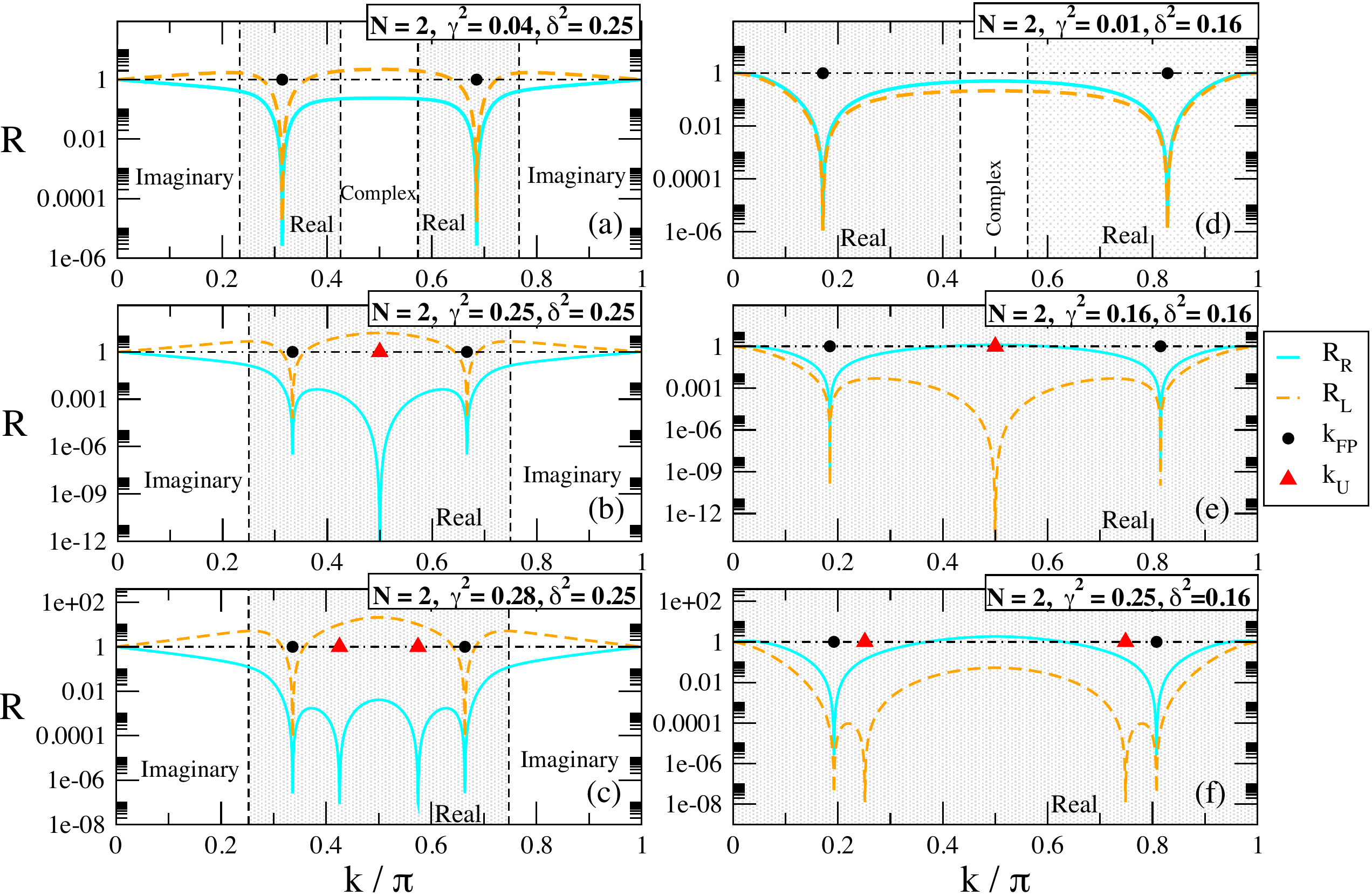} 
     \caption{\label{fig:reflex_N2}  {Left/right reflections coefficients, $R_L$ and $R_R$, as a function of the wave number $\mu$ for $N=2$ and the control parameters of Fig.~\ref{fig:mu_vs_E}. Left (right) panels correspond to $\alpha<1$ ($\alpha>1$). The Fabry-Perot resonances are marked by black full circles, while the $U$-points are shown as red triangles. Inside shaded energy regions the values of $\mu$ are real, while outside they are either imaginary or complex, according to Eq.~\eqref{eq:casos_mu}.} }
\label{fig:RR_RL_alpha>1_<1}
\end{figure}

In analogy with the data for the transmission coefficient $T$ presented in Fig.~\ref{fig:T_N2_10_y_F} and in Fig. \ref{fig:T_N2_10_y_F-}, now in Fig.~\ref{fig:reflex_N2} we show the energy dependence of $R_L$ and $R_R$ for the same model parameters. As is expected, the data in Fig.~\ref{fig:reflex_N2}(a) manifest that when the transmission is perfect, $T=1$, both $R_L$ and $R_R$ vanish (when $\gamma^{2} < \delta^{2}$). In this case, the corresponding energy values (black full circles) are defined by the Fabry-Perot resonances. However, when $\gamma^{2} > \delta^{2}$, see Fig.~\ref{fig:reflex_N2}(c), two new points (red triangles) (termed above as the $U$-points) arise, which do not correspond to the Fabry-Perot resonances. In the middle panel the data is shown for $\gamma^{2} \geq \delta^{2}$, for which the two $U$-points merge. As is mentioned before, this is the only case when {\it both} the left and right reflection coefficients equal zero. This happens at the center of the energy spectra. Note that at the $U$-points (shown in terms of the wave vector $k_{U}$) only the right reflection coefficients vanishes, while the other one, $R_L$, remains finite. This is the case of unidirectional reflectivity. 

It can be shown that at the $U$-points the functions $F_{\pm}$ take the simple form
\begin{equation}
F_{\pm} = \pm \frac{ \gamma }{\alpha} +  \frac{1}{2} \: \frac{\delta^{2} - \gamma^{2} -2 \delta }{\alpha \sin(k)} . 
\end{equation}
This expression allows us to obtain exact values for $R_L$ and $R_R$. Specifically, for $\gamma=\delta$ we get $R_R=0$ and $R_L \simeq N^{2} F_{-}^{2} \simeq 16$, while for $\gamma^{2} > \delta^{2}$ we have $R_{R} = 0$ and $
R_{L} \simeq  18$.

The $\alpha$ parameter modifies the width of the energy in accordance with the dispersion relation \eqref{eq:disp} and the change is large for the cases where $\alpha$ is greater or less than 1. We have also found that the region with imaginary energies disappears for $\alpha > 1$, see Fig.\ref{fig:mu_vs_E}. Another important effect is the interchange between left an right unidirectional reflections in an asymmetric model,  which can be observed experimentally.

\psection{Conclusion}

We have analytically developed exact expressions for the transmission and reflection coefficients for a $\mathcal{PT}$-symmetric model, valid for any values of control parameters of the model. These expressions are obtained in the form which allows one to effectively analyze all transport properties. The main interest is paid to the role played by the inclusion of the asymmetry with respect to left/right couplings between neighboring sites. We have shown that due to this asymmetry, a new kind of specific points emerge in the energy spectrum, at which the unidirectional reflectivity occurs. In contrast with the symmetric setup, in which the unidirectional reflectivity has been already observed, the asymmetric coupling results in an emergence of an internal (in the energy spectra) region where the eigenvalues have both real and imaginary parts. 

One of the specific properties of the model with asymmetric coupling is a new kind of energy values $E_{U}$ for which the vanishing of the reflection is not related to the exceptional points in the energy spectra. These $U$-points are also different from the Fabry-Perot resonances since they do not depend on the number of the dimers $N$. Instead, they are defined by the functions $F_{\pm}$ which we have introduced. We have identified the conditions, $\delta^{2} \leq \gamma^{2}$, for which these $U$-points emerge for the considered model. Our results can be confirmed experimentally by studying the wave propagation along one-mode waveguides with non-symmetric couplings.

\ack
This work was partially supported by 
VIEP-BUAP (Grant No.~MEBJ-EXC19-G),
Fondo Institucional PIFCA (Grant No.~BUAP-CA-169),
and CONACyT (Grant No.~286633).

\end{paper}

\end{document}